# Single-Walled Carbon Nanotube Sample Purification

**Saunab Ghosh, Sergei M. Bachilo, and R. Bruce Weisman**[*]

Department of Chemistry and R.E. Smalley Institute for Nanoscale Science and Technology, Rice University, 6100 Main Street, Houston, Texas 77005, USA

[*]e-mail: weisman@rice.edu

*Manuscript Text*

## INTRODUCTION

The remarkable physical and chemical properties of single-walled carbon nanotubes (SWCNTs) have motivated intense research efforts in science and engineering. A key step in most SWCNT studies or applications is the processing of raw reactor product to obtain individualized nanotubes free of the carbonaceous impurities, nanotube bundles, and residual metallic growth catalyst that can constitute substantial portions of the as-grown material.[1] It is particularly challenging to process and purify samples without causing physical or chemical damage to the nanotubes themselves. Such damage is most evident in changes to their optical properties, such as increased Raman D/G band ratios,[2] broadened or lowered-contrast absorption peaks,[3] or reduced near-IR fluorescence efficiency.[4] The first step in purification is disentangling the components of a solid raw SWCNT sample, typically by ultrasonic dispersion in an aqueous surfactant solution.[5] Nanotube damage in this step can be limited by minimizing the sonication duration and intensity. The resulting dispersions are then commonly ultracentrifuged to separate denser bundles and impurities from the individually suspended SWCNTs, which are recovered in the supernatant.[6]

We describe here a simple, inexpensive, and scalable alternative to centrifugation that uses permanent magnets to remove many impurities and bundles from nanotube suspensions. Although there are several prior reports of magnetic purification of nanotube samples,[7-10] those studies were focused only on the removal of metallic impurities. By using a range of optical and AFM measurements, we demonstrate that our magnetic processing method also removes nanotube bundles as effectively as ultracentrifugation, giving purified suspensions with sharp absorption spectra and high fluorescence efficiencies. Such samples are ideal for subsequent advanced processing by methods such as ion exchange chromatography, density gradient ultracentrifugation, or gel-based structural sorting.[11-18]

## EXPERIMENTAL METHODS

### *SWCNT sample dispersion*

HiPco SWCNTs used in this study were taken from two batches (HPR 188.4 and HPR 195) produced in the Rice University reactor.[20] Approximately 1.5 mg of raw nanotubes were added to a glass vial containing 10 mL of aqueous 1% (w/v) sodium deoxycholate (DOC) (99% purity from Fisher Scientific). The sample was ultrasonically dispersed using a Microson XL ultrasonic cell disrupter equipped with a 3 mm probe tip and set for 6 W output power. We agitated for 20 min using a sonicator duty cycle of 30 s on, 30 s off while the vial was immersed in a room temperature water bath to avoid overheating. Conventionally processed samples used for comparison with magnetic purification (see Figure 4) were further treated by centrifugation for 5 h at 40,000 rpm (171,000 g max) in a model MLS-50 rotor and Optima Max ultracentrifuge (Beckman).

## Sample characterization

### Optical spectroscopy

We measured optical absorption and near-IR fluorescence spectra using a model NS2 NanoSpectralyzer (Applied NanoFluorescence, Houston, TX). Absorption spectra of aqueous SWCNT samples in 1.0 cm path length cells were recorded relative to a matched reference from 400 to 1400 nm, with spectral resolutions of 1 nm in the visible and 4 nm in the near-IR. The spectra are displayed without background subtraction or vertical offset. Fluorescence emission spectra were measured using diode laser excitation sources at 642, 659, and 784 nm. Raman spectra of the aqueous suspensions were obtained using a model NS3 NanoSpectralyzer (Applied NanoFluorescence, Houston, TX) with 671 nm excitation.

### Near-IR fluorescence microscopy

Dilute SWCNT suspensions were drop-cast onto glass slides and mounted on the sample stage of a Nikon TE-2000U inverted microscope that had been customized, as described previously,[21] for near-IR fluorescence imaging. We used a 785 nm excitation laser and a 60x water-immersion objective lens (Nikon, PlanApo IR, NA = 1.27). To help distinguish emissions of individual and aggregated SWCNTs, image sequences were recorded with a 100 ms exposure time while the excitation beam polarization plane was rotated in 10 degree steps. We then analyzed the depth of intensity modulation caused by polarization rotation for each emissive object in the frame. Spectra of single emissive objects in the SWCNT samples were also captured by a near-IR spectrometer coupled to the microscope. These spectra were examined for the presence of single or multiple emission peaks to distinguish individual SWCNTs from aggregates.

### Atomic Force Microscopy (AFM)

To obtain AFM images of nanotubes in our samples, the DOC surfactant concentration of a suspension was decreased to 0.3 % (w/v) by diluting with water. Then a small (ca. 10 µL) portion was spin-coated onto a freshly cleaved mica surface, passively dried for 1 min, washed with 10 µL of methanol to remove excess surfactant from the mica surface, and spun for another 5 min. The AFM image was acquired over a (10 µm)$^2$ area with 512 samples/line resolution in tapping mode using a NanoScope IIIA (Digital Instruments). The height profile range was set to 5 nm. These AFM images were used to calculate the percentage of SWCNTs deposited as individual tubes from the purified suspension. SWCNTs in the image frames with measured height profiles above 1.4 nm were considered to be bundles; those below that value were classified as individuals. SWCNT lengths were measured manually from the AFM images and compiled to obtain the histogram shown in Figure S8. Objects shorter than 80 nm were not included in this analysis.

**RESULTS AND DISCUSSION**

We studied the effect of magnetic processing by comparing properties of the suspensions before and after treatment, and also by examining small samples withdrawn periodically during the processing. Characterization methods included bulk optical absorption, near-IR fluorescence, and Raman spectroscopies; TGA measurements; microscopy using AFM and near-IR fluorescence imaging; and near-IR fluorescence spectroscopy of single particles. Figure 2a shows the dramatic change in absorption spectrum between the initial sample and the supernatant obtained from magnetic processing (see Figure S1 for the same data normalized to the 983 nm peak). The broad underlying background absorption was reduced by approximately a factor of ten after full processing, while resonant features identifiable as the $E_{11}$ and $E_{22}$ transitions of specific (*n*,*m*) species were only slightly weakened, suggesting retention of most individualized nanotubes. This increase in spectral contrast is clearly seen from the plots in Figure 2b, which compare the absorbance ratio and relative height of the 983 nm peak as compared to its adjacent valley at 938 nm. SWCNT fluorescence, which is known to be emitted much more strongly from individualized nanotubes than from aggregates, is plotted in Figure 2c. Magnetic processing increased the measured fluorescence intensity by a factor of ca. 2.5. We interpret this not as an increase in the concentration of individualized SWCNTs, but instead as a decreased inner filter effect: the magnetic removal of nonemissive components such as bundles and impurities allowed more efficient penetration of the excitation and emission light through the sample, giving larger detected signals. The combined effects of reduced background absorption and stronger emission can be observed clearly in the samples' fluorescence efficiency, defined as the ratio of spectrally integrated emission to sample absorbance at the excitation wavelength.[19] Figure 2d shows that this fluorescence efficiency increases by a factor of ~25 over a 5 h magnetic purification run. We also measured Raman spectra of the SWCNT suspension using 671 nm excitation (see Figure S2). When normalized to the sample absorbance at 671 nm, the observed G-band intensity increased by a factor of ~10 during the magnetic purification as non-SWCNT impurities were steadily removed from the supernatant. We observed qualitatively similar but less dramatic optical evidence of purification for SWCNT samples grown by the CoMoCAT method, which uses the ferromagnetic metal cobalt in its catalyst (see Figure S3).

We applied Thermo Gravimetric Analysis (TGA) to analyze for the amount of Fe catalyst present in SWCNT samples before and after magnetic purification (Figure S4). Based on the residual mass values remaining after high temperature oxidation and the assumption that iron is converted into $Fe_2O_3$, we find that the mass fraction of iron is reduced from 36% to ~12% by magnetic processing. It seems likely that refinements to the processing device and protocol could provide more complete iron removal from the SWCNT dispersions.

The strong improvements in optical absorption and emission spectra shown in Figure 2 suggest that our process for removing magnetic particles also removes nanotube aggregates from the supernatant. To investigate this point, we used near-IR fluorescence microscopy to study diluted supernatant samples drop-cast onto slides. Approximately 80% of the objects observed in the microscope field showed strongly modulated emission intensity when the polarization plane of the excitation light was rotated (see Figure 3a,b). This signature of SWCNT fluorescence indicates that those emitters are either individual nanotubes or small bundles of parallel nanotubes. We also measured emission spectra of 100 randomly chosen emitters. Of these, 82 showed single-peaked spectra indicating an individual SWCNT (see Figure S5) and 18 gave more complex spectra characteristic of small aggregates (Figure S6). We supplemented the optical measurements with AFM imaging of the magnetically processed supernatant (Figure 3c and S7). As shown in Figure 3d, the distribution of measured AFM height profiles is strongly peaked at a value near 0.8 nm, as is consistent with a sample of individualized HiPco SWCNTs. Approximately 80% of the AFM-imaged nanotubes are classified as individual based on their height profiles, and the remainder appear to be mostly small bundles containing two SWCNTs. Both optical and AFM microscopy therefore indicate effective removal of nanotube aggregates and strong enrichment of individualized nanotubes in the magnetic supernatant. The average nanotube length found from AFM image analysis is approximately 1.1 μm (Figure S8). This relatively large value reflects the mild sonication conditions used for sample dispersion prior to magnetic purification.

Since 2002,[6] ultracentrifugation has served as the standard method for removing aggregates from SWCNT dispersions to enrich samples in individualized nanotubes. To compare the effectiveness of ultracentrifugation and magnetic purification, we present in Figure 4 the optical spectra of SWCNT supernatants prepared from the same starting HiPco dispersion using the two methods. Figure 4a shows that the two absorption spectra are very similar in magnitude and shape, although the nonresonant background is ~12% higher in the magnetically processed sample, indicating slightly less efficient aggregate removal. Figure 4b compares fluorescence emission spectra of the two processed samples measured under matched conditions. These two traces are almost identical. They both show emission that is much more intense than from the parent unprocessed dispersion, reflecting the removal of absorbing, nonemissive impurities. These spectral measurements, which are known to sensitively reflect SWCNT condition, aggregation state, and impurity content, confirm that magnetic purification gives samples that are comparable in quality to those prepared by lengthy ultracentrifugation.

Finally, we have examined the solid residue deposited on the permanent magnets during processing. Deposits are found to be densest along edges of the magnets, where the magnetic field gradients are largest and forces on ferromagnetic particles are expected to be greatest. When this solid residue is gently redispersed with manual agitation, it gives suspensions whose absorption spectra show large diffuse backgrounds and relatively weaker resonant features than the unprocessed sample (see Figure S9). The redispersed residue also gives fluorescence intensities that are much lower (by a factor of ~13) as compared to a processed sample with equal absorbance at 980 nm (Figure S10).

These findings support the hypothesis of a purification mechanism that disproportionately removes nanotube bundles and impurities. Presumably, a fraction of SWCNTs in the original sample dispersion remain attached to the catalytic iron nanoparticles from which they grew. These individual nanotubes and any bundles containing an iron-attached nanotube would be subject to magnetic removal. Statistically, however, a larger fraction of bundles than individuals would contain iron, and magnetic nanoparticles might also be interstitially trapped within nanotube aggregates. Magnetic processing would then preferentially remove bundles and leave supernatants enriched in individually suspended nanotubes.

## CONCLUSIONS

Using a range of bulk and single-particle characterization methods, we have demonstrated the effectiveness of a simple nanotube purification method in which a raw aqueous dispersion of HiPco or CoMoCAT SWCNTs is gently circulated over permanent magnets. The resulting supernatant is depleted not only in residual ferromagnetic catalyst, but also in bundled nanotubes, while most of the individualized SWCNTs are retained. Optical properties of these samples are very similar to those of supernatants prepared by extensive ultracentrifugation. However, the new processing method provides major advantages in equipment cost, simplicity, energy usage, and scalability. It should find widespread use in laboratory research and industrial applications.

### *Acknowledgments*

This research was supported by grants from the National Science Foundation (CHE-1112374) and the Welch Foundation (C-0807). We thank R. H. Hauge for useful discussions and X. Fan for assistance with TGA measurements.

**Supporting Information Available:** Plots of additional absorption spectral data; Raman spectra; TGA data; single particle emission spectra; AFM images; SWCNT length histogram; spectra of material removed by magnets; data showing purification of a different batch of HiPco material with lower iron content. This material is available free of charge *via* the Internet at http://pubs.acs.org.